\documentstyle[aps,floats,epsfig]{revtex}

\begin{document}

\twocolumn
\renewcommand{\topfraction}{1.0}
\twocolumn[\hsize\textwidth\columnwidth\hsize\csname
@twocolumnfalse\endcsname
\title{Full-Sky Search for Ultra High Energy Cosmic Ray Anisotropies }
\author{Luis Anchordoqui $^1$, Carlos Hojvat $^2$, Thomas McCauley $^1$,\\ Thomas~Paul $^1$, 
Stephen Reucroft $^1$,  John Swain $^1$, and Allan Widom $^1$}
\address{$^1$ Department of Physics,
Northeastern University, Boston, MA 02115\\
$^2$ Fermi National Accelerator Laboratory, 
P.O. Box 500, Batavia, IL 60510}

\maketitle

\begin{abstract}
Using data from the SUGAR and the AGASA experiments taken during a 10 yr period with nearly uniform 
exposure to the entire sky, we search for large-scale anisotropy patterns in the arrival directions of  
cosmic rays with energies~$>10^{19.6}$~eV. We determine the angular power spectrum from an
expansion in spherical harmonics for modes out to $\ell =5$. Based on available statistics, we 
find no significant deviation from isotropy. We compare the rather modest results 
which can be extracted from existing data samples with the results that should be forthcoming as 
new full-sky observatories begin operation. 
\end{abstract}

\vskip2pc]

%
%

\section{Introduction}

Ultra high energy cosmic rays are one of the most enigmatic phenomena in the 
universe. Despite the fact that the existence of particles with energies~$\agt 10^{20}$~eV has been known 
for over 40 years, their origin continues to be an intriguing puzzle~\cite{Bhattacharjee:1998qc}. 
The distribution of arrival directions is perhaps the most helpful observable in yielding clues 
about cosmic ray origin. On the one hand, if cosmic rays cluster within a small angular 
region~\cite{Uchihori:1999gu} or show directional alignment with powerful compact 
objects~\cite{Farrar:1998we}, one might be able to associate them with isolated sources in the sky.  
On the other hand, if the distribution of arrival directions exhibits a large-scale anisotropy, this could 
indicate whether or not certain classes of sources are associated with large-scale structures (such 
as the Galactic plane or the Galactic halo). In this paper, we focus our attention on the search for 
such large-scale celestial patterns.

Cosmic ray air shower detectors which experience stable operation over a
period of a year or more will have a uniform exposure in right ascension,
$\alpha$. A traditional technique to 
search for large-scale anisotropies is then to fit the right ascension distribution of events to a sine 
wave with period $2\pi/m$ ($m^{\rm th}$ harmonic) to determine the components ($x, y$) of the Rayleigh 
vector~\cite{Linsley}
\begin{equation}
x = \frac{2}{N} \sum_{i=1}^{N} \, \cos(m\, \alpha_i) \,, \,\,\,\,\,y = \frac{2}{N} \sum_{i=1}^{N} \,\, 
\, \sin( m\, \alpha_i)\,.
\end{equation}
The $m^{\rm th}$ harmonic amplitude of $N$ measurements $\alpha_i$ is given by the Rayleigh vector 
length ${\cal R}~=~(x^2~+~y^2)^{1/2}$. The expected length of such vector for values randomly 
sampled from a uniform phase distribution is ${\cal R}_0~=~2/\sqrt{N}$.  The chance probability 
of obtaining an amplitude with length larger than that measured is
$p(\geq~{\cal R})~=~e^{-k_0},$ where $k_0~=~{\cal R}^2/{\cal R}_0^2.$ 
To give a specific example, a vector of length $k_0~\geq~6.6$ would be required to 
claim an observation whose probability of arising from random fluctuation
was 0.0013 (a ``$3\sigma$'' result). For the ultra high energy~($\agt~10^{19.6}$~eV) regime, all 
experiments to date have reported $k_0 \ll 6.6$~\cite{Edge:rr}. This does not imply an isotropic 
distribution, but it merely means that available data are too sparse to claim a statistically significant 
measurement of anisotropy by any of these experiments.
In other words, there may exist anisotropies at a level too low to 
discern given existing statistics~\cite{Evans:2001rv}. 

The right harmonic analyses are completely blind to intensity variations which depend only on declination, 
$\delta$.  Combining anisotropy searches in $\alpha$ over a range of declinations 
could dilute the results, since significant but out of phase Rayleigh vectors from different 
declination bands can cancel each other out.  Moreover, the analysis methods that consider
distributions in one celestial coordinate, while integrating
away the second, have proved to be potentially misleading~\cite{Wdowczyk:rb}. An unambiguous interpretation 
of anisotropy data requires two ingredients: {\it exposure to the full celestial sphere and 
analysis in terms of both celestial coordinates.} Though the 
statistics are very limited at present, this article describes  
a first step in this direction. In the next section we combine data 
from the Sydney University Giant Air-shower Recorder (SUGAR) and the Akeno Giant Air Shower Array (AGASA) 
taken during a 10 yr period with nearly uniform exposure to the entire sky. After that, in Sec.~III, we 
apply the power spectrum estimation technique~\cite{Peebles} to interpret the distribution of arrival 
directions. Our conclusions are collected in Sec.~IV.

\section{Experimental data sets}

The SUGAR array was operated from January 1968 to February 1979 in  New South Wales (Australia) at a 
latitude of $30.5^\circ$ South and longitude $149^\circ 38'$ East~\cite{Winn:un}. The array consisted 
of 47 independent  
stations on a rectangular grid covering an area $S \approx 70$~km$^2$. The primary energy was determined 
from the total number of muons, $N_\mu$, traversing the detector at the measured zenith angle $\theta$. 
The total aperture for incident zenith angles between $\theta_1$ and $\theta_2$ was found to be
\begin{equation}
A =  \int_{\theta_1}^{\theta_2} S \,\,p(N_\mu, \theta) \, \cos \theta \, d\Omega \,.
\end{equation}
Here, $p(N_\mu, \theta)$ is the probability that a shower falling within the physical area
was detected, $S \cos \theta$ is the projected surface of the array in the shower plane, and 
$d\Omega$ is the acceptance solid angle. The SUGAR Collaboration reports~\cite{Winn:un}
a reasonable accuracy in assessing the shower parameters up to $\theta = 73^\circ$. The particulars of the events with primary energy $> 10^{19.6}$~eV are 
given in Table~I. 
The estimated angular uncertainty for showers that triggered 5 or more stations is reported 
as $3^\circ \sec \theta$~\cite{Winn:un}. However, the 
majority of events were only viewed by 3 or 4 stations, 
and for these the resolution appears to be as poor as $10^\circ$~\cite{Kewley:zt}. Of particular 
interest for this analysis,
\mbox{$p(N_\mu > 10^8, \theta < 55^\circ) \approx 0.85$}~\cite{Bell:gp}, yielding a total aperture $A
\approx 125$~km$^2$~sr. This provides an exposure reasonably matched to that of AGASA, which is 
described next.

The AGASA experiment occupies farm land near the village of Akeno (Japan) at a longitude 
of $138^\circ 30'$ East and  latitude $35^\circ 30'$ North~\cite{Chiba:1991nf}. The array, 
which consists of 111 
surface detectors deployed over an area of about 100 km$^2$, has been running since 1990. 
About 95\% of the surface detectors were operational from March to December 1991, and
the array has been fully operational since then.  A prototype detector operated from 1984 to 1990  
and has been part of AGASA since 1990~\cite{Teshima:1985vs}.
The aperture for events with primary zenith angle $0^\circ < \theta < 45^\circ$ and energies 
beyond $10^{19.25}$~eV is found to be $A \approx 125$~km$^2$ sr~\cite{Chiba:1991nf}. 
The angular resolution for these events is $1.6^\circ$~\cite{Takeda:1998ps}. 
The arrival directions of cosmic rays with energy $> 10^{19.6}$~eV are given 
in Table~II.

The expected 
event rate is found to be
\begin{eqnarray}
\frac{dN}{dt} &  = &  A\, \int_{E_1}^{E_2}\, E^3 J(E)\, \frac{dE}{E^3} \nonumber \\ 
 & \approx & \frac{A}{2} \,\langle E^3\, J(E) \rangle\, \left[ \frac{1}{E_1^2} - \frac{1}{E_2^2} \right] \,\,,\label{flux}
\end{eqnarray}
where $\langle E^3 J(E) \rangle \approx 10^{24.6}$~eV$^2$ m$^{-2}$ s$^{-1}$ sr$^{-1}$ stands for the 
observed ultra high energy cosmic ray flux, which has a cutoff at $E_2 = 10^{20.5}$~eV~[1]. Thus, 
in approximately 10~yr of running each of 
these experiments should collect $\approx 50$ events above $E_1 = 10^{19.6}$~eV, arriving with a zenith 
angle $< \theta_{\rm max}$. Here, $\theta_{\rm max} = 45^\circ$ 
for AGASA and $\theta_{\rm max} = 55^\circ$ for SUGAR. Our sub-sample for the full-sky anisotropy 
search consists of the 50 events detected by AGASA from May 1990 to May 2000~[6], 
and the 49 events detected by SUGAR  with $\theta < 55^\circ$~[11]. Note that we consider the full data sample for the 11 yr lifetime of SUGAR (in contrast to the 10 yr data sample from AGASA). This roughly compensates for the time variation of the sensitive area of the experiment as detectors were deployed or inactivated for 
maintenance. The arrival directions of the 99 events are plotted in Fig.~1 (equatorial coordinates 
B.1950).

\begin{figure}
\label{fig2}
\begin{center}
\epsfig{file=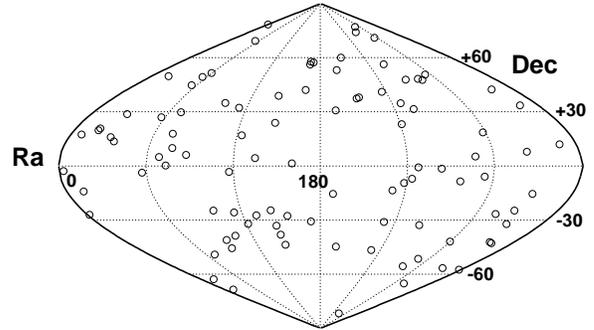,width=8.cm,clip=} 
\caption{Arrival direction of the 99 events observed above $10^{19.6}$~eV by the SUGAR ($\theta < 55^\circ$) 
and the AGASA ($\theta < 45^\circ$) experiments (equatorial coordinates 
B.1950).}
\end{center}
\end{figure}

A detector at latitude $a_0$ that has continuous operation with constant 
exposure in right ascension and is fully efficient for $\theta < \theta_{\rm max}$ has relative exposure
with the following dependence on declination~\cite{Sommers:2000us} 
\begin{equation}
\omega (\delta) \propto (\cos a_0\,\,\cos\delta\,\,\sin\alpha_{\rm max} + \alpha_{\rm max} 
\,\,\sin a_0\, \,\sin \delta)\,\,,
\label{omeguita}
\end{equation}
where $\alpha_{\rm max}$, the local hour angle at which the zenith angle becomes 
equal to $\theta_{\rm max}$, is given by
\begin{equation}
\alpha_{\rm max} = \left\{ \begin{array}{ll}
0 & {\rm if}\,\,\,\xi > 1 \\
\pi & {\rm if} \,\,\, \xi < -1 \\
\cos^{-1}\,\,\xi & {\rm otherwise}
\end{array} \right.
\end{equation}
with
\begin{equation}
\xi \equiv \frac{\cos \theta_{\rm max} - \sin a_0\,\,\sin \delta}{\cos a_0\,\,\,\cos\delta}\,\,.
\end{equation} 
We normalize the relative exposure by scaling the declination angle 
dependence given in Eq.~(\ref{omeguita}) by an estimate of the areas and 
operation times for SUGAR and AGASA (recall that 
the stated area of SUGAR is an overestimate, and this compensated to some degree by the longer running time).   
The result, displayed in Fig.~2, shows that the combined exposure of these 
arrays is nearly uniform over the entire sky.

\begin{figure}
\label{fig1}
\begin{center}
\epsfig{file=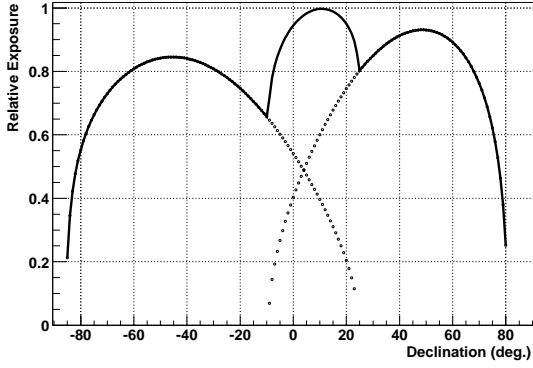,width=8.cm,clip=} 
\caption{Declination dependence of SUGAR and AGASA relative exposures (dotted). The solid line indicates  
the combined declination dependence of the relative exposure.}
\end{center}
\end{figure}

\section{Correlations and Power Spectrum}

We begin this section with a general introduction to
the calculation of the angular power spectrum and the determination of
the expected size of intensity fluctuations.  The technique is then
applied to the AGASA and SUGAR data in order to check for
fluctuations beyond those expected from an isotropic distribution.

Let us start by defining the directional phase space of the
angular distribution of cosmic ray events in equatorial coordinates, $(\alpha, \delta)$. 
(i) The direction of the event is described by a unit vector
\begin{equation}
{\bf n}=\sin\delta \,({\bf i}\,\cos\alpha +{\bf j}\, \sin\alpha )+{\bf k}\,\cos\delta \,\,;
\label{CP1}
\end{equation}
(ii) The solid angle is given by
\begin{equation}
d^2{\bf n}=\sin\delta \,\,d\delta \,d\alpha \,\,;
\label{CP2}
\end{equation}
(iii) The delta function for the solid angle is defined as
\begin{equation}
\delta ({\bf n},{\bf n}^\prime ) =
\delta (\cos \delta -\cos \delta^\prime )\,
\sum_{m=-\infty }^\infty
\delta (\alpha -\alpha^\prime +2\pi m)\,,
\end{equation}
so that, as usual,
\begin{equation}
f({\bf n}) = \int
\delta ({\bf n},{\bf n}^\prime )\,\,f({\bf n}^\prime )\,\,d^2{\bf n}^\prime \,\,;
\label{CP3}
\end{equation}
(iv) The probability distribution
\begin{math} P({\bf n})d^2{\bf n} \end{math}
of events can be employed for the purpose of computing the averages
\begin{equation}
\overline{f}=\int f({\bf n})\,\,P({\bf n})\,\,d^2{\bf n}\,\,;
\label{CP4}
\end{equation}
Finally, (v) for a sequence of \begin{math} N \end{math} different
cosmic ray events \begin{math}({\bf n}_1,\ldots ,{\bf n}_N)\end{math}
one may assume an independent distributions for each event, i.e.
\begin{equation}
P_N({\bf n}_1,\ldots ,{\bf n}_N)\prod_i^N d^2{\bf n}_i=
\prod_i^N \{P({\bf n}_i)\,d^2{\bf n}_i\}\,\,.
\label{CP5}
\end{equation}

For a sequence of events
\begin{math}({\bf n}_1,\ldots ,{\bf n}_N)\end{math}
let us describe the angular intensity as
the random variable
\begin{equation}
I({\bf n})=\frac{1}{ N}\sum_{j=1}^N \,  \delta ({\bf n},{\bf n}_j)\,\,.
\label{CP6}
\end{equation}
From Eqs.~(\ref{CP5}) and (\ref{CP6}) it follows that
\begin{eqnarray}
\overline{I({\bf n})} & = &
\int \ldots \int
I({\bf n})\,P_N({\bf n}_1,\ldots ,{\bf n}_N)\,\prod_i^N d^2{\bf n}_i \nonumber \\
&  = & P({\bf n}).
\label{CP7}
\end{eqnarray}
The two point correlation function
\begin{math} G({\bf n},{\bf n}^\prime )
=\overline{I({\bf n})I({\bf n}^\prime )} \end{math}
is defined via
\begin{eqnarray}
G({\bf n},{\bf n}^\prime )&=&\int \ldots \int
I({\bf n})\ I({\bf n}^\prime ) \
P_N({\bf n}_1,\ldots ,{\bf n}_N) \ \prod_i^N d^2{\bf n}_i \nonumber \\
 &=& \frac{1}{N}\,\delta ({\bf n},{\bf n}^\prime )\,P({\bf n})+
\left(1-\frac{1}{N}\right)\,P({\bf n})\,P({\bf n}^\prime ) \,\,. \nonumber\\
 & & 
\label{CP8}
\end{eqnarray}

The ``power spectrum'' of the correlation function is determined by
the eigenvalue equation
\begin{equation}
\int G({\bf n},{\bf n}^\prime )\,\,\psi_\lambda ({\bf n}^\prime )\,\,
d^2{\bf n}^\prime=\lambda \,\,\psi_\lambda ({\bf n}).
\label{CP9}
\end{equation}
In this regard it is useful to introduce Dirac notation to indicate the inner product
\begin{equation}
\left<\psi|\psi\right>=
\int \psi^*({\bf n})\,\,\psi({\bf n})\,\,d^2{\bf n}\,\,.
\label{CP10}
\end{equation}
With this in mind, Eq.~(\ref{CP9}) reads
\begin{equation}
G \,\left|\psi_\lambda \right>=\lambda \,\left|\psi_\lambda \right>.
\label{CP11}
\end{equation}

In the limit of a large number of events
\begin{math} N\to \infty  \end{math},
\begin{equation}
\lim_{N\to \infty}G({\bf n},{\bf n}^\prime )
\equiv G_\infty ({\bf n},{\bf n}^\prime )
=P({\bf n})P({\bf n}^\prime )\,,
\end{equation}
or equivalently,
\begin{equation}
\hat{G}_\infty = \left|P\right>\left<P\right| \,.
\label{CP12}
\end{equation}
In such a limit, fluctuations can be neglected and we
find only {\em two
possible values} in the spectrum: (i) There is a
non-degenerate non-zero
eigenvalue
\begin{equation}
\hat{G}_\infty \left|P\right> = \lambda_\infty
\left|P\right>\,,
\end{equation}
with
\begin{equation}
\lambda_\infty = \left<P|P\right> = \int P^2({\bf
n})d^2 {\bf n}.
\label{CP13}
\end{equation}
(ii) For every state \begin{math} \left| f \right>
\end{math}
orthogonal to \begin{math} \left| P \right>
\end{math}
with mean value \begin{math}
\bar{f}=\left<P|f\right>=0 \end{math},
there exists a {\em zero eigenvalue} in the power
spectrum
\begin{equation}
\hat{G}_\infty \left|f\right> = 
\left|P\right>\left<P|f\right>
=\left\{\int P f d^2{\bf n}\right\}\left|P\right>
=\bar{f}\left|P\right> = 0 \,\,.
\label{CP14}
\end{equation}
Let us  now turn  to consider the effects of finite \begin{math}
N \end{math}. Defining the fluctuations in the intensity by
\begin{equation}
\Delta I({\bf n})=I({\bf n})-\overline{I({\bf n})}
=I({\bf n})-P({\bf n}),
\label{CP15}
\end{equation}
the two point correlation function can be re-written as
\begin{eqnarray}
G({\bf n},{\bf n}^\prime ) &  = &
\overline{I({\bf n})I({\bf n}^\prime )} =
\overline{I({\bf n})}\ \overline{I({\bf n}^\prime )}
+\overline{\Delta I({\bf n})\Delta I({\bf n}^\prime )}
\nonumber \\
 & = & G_\infty ({\bf n},{\bf
n}^\prime )
+\overline{\Delta I({\bf n})\Delta I({\bf n}^\prime )}\,,
\end{eqnarray}
with
\begin{equation}
\overline{\Delta I({\bf n})\Delta I({\bf n}^\prime )}
 = \frac{1}{N}\,\left[\, \delta ({\bf n},{\bf n}^\prime
)P({\bf n})
-P({\bf n})P({\bf n}^\prime )\,\right] \,,
\label{CP16}
\end{equation}
where Eq.~(\ref{CP8}) has been invoked. Putting all this together, some general results 
follow: (i) For the \mbox{$N\to\infty$} case, there is
only one state
with a finite eigenvalue \begin{math} \lambda_\infty
\end{math}, while
the rest of the power spectrum corresponds to
\begin{math} \lambda =0  \end{math}. (ii) For finite
\begin{math} N \end{math}, Eq.~(\ref{CP16}) implies that
the fluctuations are of order
\begin{math} N^{-1} \end{math}. The power
spectrum for large \begin{math} N \end{math} then has
one eigenvalue of order unity and the rest of the eigenvalues are of
order
\begin{math} N^{-1} \end{math}.

\begin{figure}
\label{fig4}
\begin{center}
\epsfig{file=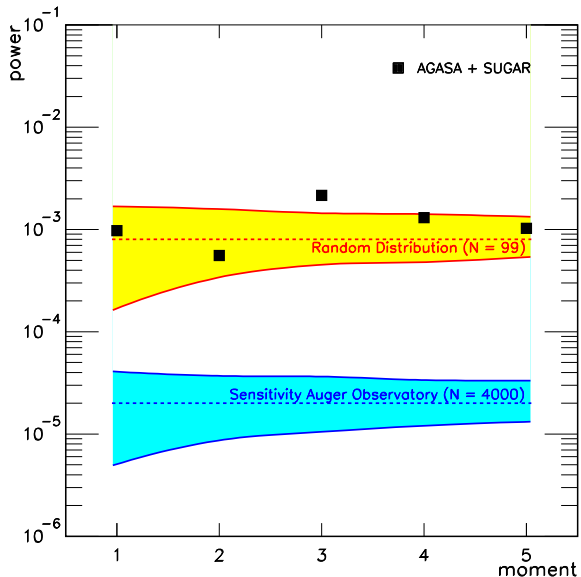,width=8.cm,clip=} 
\caption{The angular power spectrum is indicated by the squares.  The horizontal lines
indicate the mean value, $\overline{C}(\ell) = (4 \pi N)^{-1},$ expected for an isotropic 
distribution. The upper shaded band shows the 1$\sigma$ fluctuation around the mean value for
$N=99$. The band was obtained from 1000 sets of Monte Carlo simulations
of 99 events each, including small corrections for $\omega_j$.  The resuts 
from the Monte Carlo are consistent with the errors computed using Eq.~(\ref{A14})
for the case of $\omega_j=1$.
For $\ell=3$, where there is a 
small excess compared to the expectation for isotropy, $C(3) = 2.16\times 10^{-3}$ while 
the expectation from a random distribution  is $\overline{C}_{\rm MC} (3) = 9.5 \times 10^{-4}$, 
with a variance of $5.0 \times 10^{-4}$. The projected sensitivity for the Pierre Auger Observatory  
is also indicated on the plot by the lower shaded band.  
}
\end{center}
\end{figure}

Now, for an isotropic distribution of ${\bf n}$,
\begin{equation}
\widetilde{P} ({\bf n}) = \frac{1}{4\pi } \,\,,
\end{equation}
and the two point correlation function of Eq.~(\ref{CP8}) becomes,
 \begin{equation}
 \widetilde{G} ({\bf n},{\bf n}^\prime ) =
 \frac{1}{4\pi N}\,\,\delta ({\bf n},{\bf n}^\prime )+
 \frac{1}{(4\pi )^2}\,\left({1-\frac{1}{N}}\right) \,\,.
 \label{CP17}
 \end{equation}
 The eigenvalue problem is solved by employing spherical
 harmonics~\cite{ylm} 
 \begin{equation}
 \int \widetilde{G} ({\bf n},{\bf n}^\prime ) \,\,
 Y_{lm}({\bf n}^\prime )\,
 d^2{\bf n}^\prime= \lambda_{\ell m}\,\, Y_{\ell m}({\bf n})\,\,,
 \label{CP18}
 \end{equation}
 where from Eq.~(\ref{CP17}) we have,
\begin{equation}
\lambda_{\ell m} = \left\{ \begin{array}{ll}
(4\pi)^{-1} & \hspace{1cm} \textrm{if $(\ell, m) = (0,0)$} \\
(4\pi N)^{-1} & \hspace{1cm} \textrm{if $(\ell, m) \ne (0,0)$}
\end{array}
\right.\,.
\label{CP19}
\end{equation}
The eigenfunctions form a useful set for expansions of the intensity over the celestial sphere
\begin{equation}
I({\bf n})= \sum_{\ell = 0}^\infty \,\, \sum_{m = -\ell}^\ell\,\, a_{\ell m}\,  Y_{\ell m}({\bf n})\,\, .
\label{CP20}
\end{equation}
To incorporate the dependence on declination given in Eq.~(\ref{omeguita}),
let us re-define the angular intensity
\begin{equation}
I({\bf{n}}) = \frac{1}{\cal N}\,\,\sum_{j = 1}^N  \frac{1}{\omega_j} \,\, \delta ({\bf n}, {\bf n}_j) \,\,, 
\label{I}
\end{equation}
where $\omega_j$ is the relative exposure at arrival direction ${\bf n}_j$ given in Fig.~2 and ${\cal N}$ 
is the sum of the weights $\omega_j^{-1}$. 
Since the eigenvalues of the $Y_{\ell m}$ expansion are 
uniquely defined
\begin{equation}
a_{\ell m} = \int  I ({\bf n}) \,\, Y_{\ell m} ({\bf n}) \,\, d^2{\bf n}\,\,,
\label{aintegral}
\end{equation}
the replacement of Eq.~(\ref{I}) into Eq.~(\ref{aintegral}) leads to the explicit form 
of the coefficients for our set of arrival directions 
\begin{equation}
a_{\ell m}=\frac{1}{{\cal N}}\sum_{j=1}^N \, \frac{1}{\omega_j} \,\, Y_{\ell m}({\bf n}_j) \,.
\label{CP21}
\end{equation}
Squaring Eq.~(\ref{CP21}) and taking the average yields
\begin{equation}
\overline{a_{\ell m}^2}= \frac{\delta_{\ell 0} \, \delta_{m 0}}{4\pi} + 
\frac{1}{4\pi N} \, (1 - \delta_{\ell 0} \, \delta_{m 0}) \ .
\end{equation}
Equivalently, the mean square fluctuations of the coefficients are determined 
by the power spectrum eigenvalues according to
\begin{equation}
\overline{a_{\ell m}^2}=\lambda_{\ell m}.
\label{CP22}
\end{equation}
 Although 
full anisotropy information is encoded into the coefficients $a_{\ell m}$ (tied to some 
specified coordinate system), the (coordinate independent) total power spectrum of fluctuations 
\begin{equation}
C(\ell) = \frac{1}{(2 \ell +1)}\,\, \sum_{m=-\ell}^\ell a_{\ell m}^2\,\,,
\label{CP23}
\end{equation}
provides a gross summary of the features present in the celestial distribution together with the 
characteristic angular scale(s). Note that Eqs.~(\ref{CP19}) and (\ref{CP22}) imply
\begin{equation}
\overline{C}(\ell) =  \frac{1}{(2 \ell +1)}\,\, \sum_{m=-\ell}^\ell \overline{a_{\ell m}^2}\,=
\left\{ \begin{array}{ll}
(4\pi)^{-1} &  \textrm{if $\ell = 0$} \\
(4\pi N)^{-1} & \textrm{if $\ell  \ne 0$}
\end{array}
\right.\,.
\label{Cp24}
\end{equation}

One must take care when combining data from experiments
with differing angular resolutions, as in the case for SUGAR and AGASA.
In this analysis, we consider only anisotropy on an angular scale larger
than the the resolution of SUGAR, which has the worse resolution of the two. 
The power in mode $\ell$ is sensitive to variation over 
angular scales of  $\ell^{-1}$ radians~\cite{Sommers:2000us}. 
Recalling that the 
estimated angular uncertainty for some of the events in 
the SUGAR sample is possibly as poor as $10^\circ$~\cite{Kewley:zt} we only look in this 
study for large scale patterns, going into the multipole expansion out to $\ell =5$.

Our results at this juncture are summarized in Fig.~3. The angular power spectrum is 
consistent with that expected from a random distribution for all (analyzed) multipoles, though there is a 
small ($2\sigma$) excess in the data for $\ell =3$. The majority of this excess comes from SUGAR 
data~\cite{Isola:2002ei}. 
The decrease in error as $\ell$ increases may be understood as
a consequence of the fact that contributions to mode $\ell$
arise from variations over an angular scale $\ell^{-1}$.
If one compares to the expectation for isotropy, structures
characterized by a smaller angular scale, and hence
larger $\ell$, can be ruled out with
more significance than larger structures. 

To quantify the error, we study the fluctuations in $C(\ell)$ for $\ell \geq 1$. 
For simplicity, let us neglect the small effects of declination 
({\it viz.,} $\omega_j = 1 \,\,\forall j$), and consider the random variable
\begin{equation}
X_{\ell}=\frac{C(\ell)}{\overline{C}(\ell)}=
\left(\frac{4\pi N}{2\ell +1}\right)\sum_{m=-\ell}^\ell a_{\ell m}^2.
\label{A4}
\end{equation}
Denoting by \begin{math} P_\ell(\cos \delta)  \end{math} the Legendre polynomial
of order  \begin{math} \ell \end{math} and employing the addition theorem for
spherical harmonics,
\begin{equation}
\frac{4\pi }{2\ell+1}\sum_{m=-\ell}^\ell Y_{\ell m}({\bf n})Y_{\ell m}({\bf n}^\prime )
=P_\ell ({\bf n\cdot n}^\prime) \,,
\label{A5}
\end{equation}
Eqs.~(\ref{CP21}), (\ref{A4}), and (\ref{A5}) imply that
\begin{equation}
X_\ell=1+\frac{2}{N}\sum_{1\le i<j\le N}P_\ell({\bf n}_i\cdot {\bf n}_j)\,,
\label{A6}
\end{equation}
where $P_\ell (1) = 1$ has been invoked.
Evidently, $\overline{X_\ell}=1$. Besides,
\begin{equation}
\overline{X_\ell^2}=1+\frac{4}{N^2}\sum_{1\le i<j\le N}\ \sum_{1\le k<q\le N}
\overline{P_\ell({\bf n}_i\cdot {\bf n}_j)P_\ell({\bf n}_k\cdot {\bf n}_q)}\,.
\label{A8}
\end{equation}
Since different pairs in the sum on the right hand side of Eq.~(\ref{A8})
are uncorrelated for an isotropic distribution, it follows that
\begin{equation}
\overline{X_\ell^2}=1+\frac{4}{N^2}\sum_{1\le i<j\le N}
\overline{P_\ell({\bf n}_i\cdot {\bf n}_j)^2} \,.
\label{A9}
\end{equation}
There are \begin{math} \{N(N-1)/2\} \end{math} equivalent pairs in
Eq.~(\ref{A9}) which implies
\begin{equation}
\overline{X_\ell^2}= \overline{X_\ell}^2+2\left(1-\frac{1}{N}\right)
\overline{P_\ell({\bf n}_1\cdot {\bf n}_2)^2}\,.
\label{A10}
\end{equation}
>From Eq.~(\ref{A5}) we obtain
\begin{eqnarray}
\overline{P_\ell({\bf n}_1\cdot {\bf n}_2)^2} & = &
\left(\frac{4\pi}{2\ell+1}\right)^2\sum_{m=-l}^l
\overline{Y_{\ell m}({\bf n}_1)^2}\ \overline{Y_{\ell m}({\bf n}_2)^2} \nonumber \\
 & = & \frac{1}{2\ell+1} \,.
\label{A11}
\end{eqnarray}
Plugging Eq.~(\ref{A11}) into Eq.~(\ref{A10}) leads to
\begin{equation}
\overline{\Delta X_\ell^2}=\left(1-\frac{1}{N}\right)\frac{2}{2\ell+1},
\label{A12}
\end{equation}
or equivalently, 
\begin{equation}
\left(\frac{\overline{\Delta C_\ell}^2}{\overline{C_\ell}^2}\right)
=\left(1-\frac{1}{N}\right)\frac{2}{2\ell+1} \,,
\label{A13}
\end{equation}
yielding (for large \begin{math} N \end{math})
\begin{equation}
\lim_{N\to \infty }
\left(\frac{\overline{\Delta C_\ell}^2}{\overline{C_\ell}^2}\right)
=\frac{2}{2\ell+1}\ \ {\rm for}\ \ \ell \ge 1\,,
\label{A14}
\end{equation}
which is the variance on $X_\ell$.

\section{Outlook}

We have made a first full-sky anisotropy search using
data from the SUGAR and AGASA experiments.  At present, low statistics
and poor angular resolution limits our ability to perform a very sensitive
survey, but we can at least have a preliminary look at the first moments in the
angular power spectrum. The data are consistent with isotropy, though
there appears to be a small excess for $\ell=3$, arising mostly from the
SUGAR data.

There are two caveats in this analysis which should be kept in mind.
First, from the published SUGAR results, it is difficult to make an
exact determination of the exposure, as the sensitive area of the experiment
varied as a function of time.  Here, we assumed an area-time product of
approximately 775~km$^2$ yr.  Second, there is some uncertainty in
the energy calibration.  The SUGAR results are reported in terms of the
number of vertical equivalent muons together with two possible models to
convert this to primary energy.  We have chosen the model yielding an energy
spectrum which is in better agreement with the AGASA results~\cite{Anchordoqui:2003zk}.  
It should be noted, though, that this spectrum does not agree well with the results
of the High Resolution Fly's Eye (HiRes) experiment~\cite{Abu-Zayyad:2002ta}. 
Though there are uncertainties in 
the energy scale, the impact on this anisotropy search may not be so severe.  This is because
the energy cut of $10^{19.6}$~eV is well above the last break in the spectral
index at $\sim 10^{18.5} - 10^{19}$~eV, and one would expect that all cosmic rays
above this break share similar origins.

In the near future we expect dramatically superior results from
the Pierre Auger Observatory.  This observatory is designed to measure the energy and arrival direction 
of ultra high energy cosmic rays with unprecedented precision. It will consist 
of two sites, one in the Northern hemisphere and one in the Southern, each covering an area 
$S \approx 3000$~km$^2$. The Southern site is currently
under construction while the Northern site is pending.
Once complete, these two sites together
will provide the full sky coverage and well matched exposures
which are crucial for anisotropy analyses. The base-line design of the detector includes a ground array 
consisting of 
1600 water \v{C}erenkov detectors overlooked by 4 fluorescence eyes. The angular and energy resolutions 
of the ground arrays are typically less than $1.5^\circ$ (multi-pole expansion $\ell \sim 60$) 
and less than 20\%, respectively. The detectors are designed to be fully 
efficient ($p \approx 1$) out to $\theta_{\rm max} = 60^\circ$ beyond $10^{19}$~eV, yielding a 
nearly uniform sky $A \approx 1.4 \times 10^4$~km$^2$~sr~\cite{Sommers:2000us}. In 10~yr of running the 
two  arrays will collect $\approx 4000$ events 
above $E_1 = 10^{19.6}$~eV. As can be seen from Fig.~3, such statistics will allow one to 
discern asymmetries at the level of about 1 in $10^4$.

\section*{Acknowledgments}
This work has been partially supported by 
the US National Science Foundation (NSF), under grant No. PHY-0140407.

\onecolumn
\begin{table}
\caption{Arrival directions of cosmic rays with energy $> 10^{19.6}$~eV 
in equatorial coordinates ($\alpha,\delta$) together with the incident zenith angle ($\theta$) 
as reported by the SUGAR Collaboration~\protect\cite{Winn:un}.} 
\begin{tabular}{|ccc|ccc|ccc|}
$\theta$ [deg.] & $\alpha$ [deg.] & $\delta$ [deg.] &  $\theta$ [deg.] & $\alpha$ [deg.] 
& $\delta$ [deg.]& $\theta$ [deg.] & $\alpha$ [deg.] & $\delta$ [deg.]  \\
\hline
\hline
70 &  187.5 &  31.8 & 27 &  333.0  &  -56.6 &
72 &  354.0 & -75.2 \\ 28 &  117.0  &   -3.3 &
54 &  231.0 & -31.0 &  43 &  357.0  &  -57.4 \\
43 &  147.0 & -43.6 & 34 &  130.5  &  -27.4 & 
24 &  121.5 & -32.0 \\ 31 &  105.0  &  -38.7 &
23 &  288.0 & -51.4 & 64 &   204.0 &  24.0 \\
29 &  231.0 & -13.5 & 28 &    57.0 & -3.7 &
35 &  261.0 & -32.8 \\ 38 &  135.0  &  4.3 &
60 &  340.5 &  -43.2& 51 &  300.0 &   -41.9 \\
32 &  238.5 &  -9.5 & 38 &  16.5 &  -68.5 &
68 &  181.5 &  20.1 \\ 31 &  247.5&   -0.7 &
33 &  331.5 &  -32.2 & 23 &  69 &  -49.1 \\
40 &  280.5 &  -55.6 & 64&   303&   13.6 &
62 &  126.0 &  -39.9 \\ 55&   226.5 & -31.1 &
42 &  316.5 &  -65.1 & 45 &  189 &  -15.5 \\
6 & 114 &  -25.8  &  14 &  145.5 &  -38.1 &
48 & 154.5 & -27.7 \\ 53 & 268.5 & -81.7 &
69 & 216 &  18.6 & 46 &  76.5 &  10 \\
69 & 315 & 0.7 &67 & 154.5 & 3.1 &
67 & 123 & -38 \\ 23 &  277.5 &  -8.5 &
38 & 160.5 &  1.5 & 53 &  1.5 & -27.1 \\
54 &  337.5 &  -42.2 &  57 &  85.5 &  20.5 &
70 &  88.5 &  -64.6 \\ 64 &  303 &  -47.2 &
44 & 142.5 & -24.6 & 52 &  172.5 &  -30.8 \\
15 &  99 &  -24.7 & 49 &  30 &  15.9 &
49 &  73.5 &  0.2 \\ 58 & 355.5 &  -0.5 & 
65 & 283.5 &  -50.6 & 42 &  19.5 &  -62.8 \\ 
27 & 331.5 & -15.5 & 15 & 144 &  -33.1 &
41 & 12  & -14.1 \\ 68 & 163.5 &  -5.9 &
63 & 27 & 3.1 & 46 &  315 &  -26.6 \\
56 & 234 &  1.2 & 43 &  94.5 &  -41 & 
57 & 229.5 &  -78.7 \\ 72 &  117 &  -35.9 &
62 & 262.5 &  28.2 & 22 & 93 & -45.6 \\
59 & 223.5 &  -88.3 & 40 &  3 & -2.9 &
33 &  195 &  -44.7 \\ 59 &  30 & -78.5 &
38 & 327 &  -24.7 & 23 & 231 & -46.6 \\
52 & 340.5 & -42.8& 41 &  87 &  6.1 &
58 & 183 & 4.5 \\ 57 & 22.5 &  22.2 &
61 &  244.5 &  14.3 & 65 &  255 &  -83 \\
57 & 88.5 &  -47.8 & 57 & 180 & -22.7 & & & \\
\end{tabular}
\label{tt}
\end{table}

\begin{table}
\caption{AGASA events with $\theta < 45^\circ$ and mean energy $> 10^{19.6}$~eV 
in equatorial coordinates ($\alpha, \delta$) based on the B.1950 reference frame. 
The first 2 rows refer to events measured using the prototype 
detector before  1990~\protect\cite{Hayashida:2000zr}.}
\begin{tabular}{|cc|cc|cc|cc|}
$\alpha$ [deg.] & $\delta$ [deg.] & $\alpha$ [deg.] & $\delta$ [deg.] & 
$\alpha$ [deg.] & $\delta$ [deg.] & $\alpha$ [deg.] & $\delta$ [deg.] \\
\hline
\hline
334.70  &  38.15 & 276.80  &  35.26 & 68.71 & 30.00 & 210.03  &  50.14\\
328.68  &  27.36 & 206.45  &  34.95 &  86.65  & 58.48 & & \\
\hline
243.58  &  -7.08 & 236.31  &  41.15 & 284.40  &  47.73&  53.01  &  69.33\\
287.03  &  77.12& 2.29    &  78.32 & 143.23  &  38.82 & 267.68  &  47.91\\
7.85    &  17.42 & 255.27  &  31.47 & 171.55  &  57.37 &100.17  &  34.95\\
123.54  &  16.95 & 208.34  &  60.04 & 28.28   &  28.75 &18.25   &  49.74\\
18.07   &  20.83 & 280.91  &  48.25 & 73.27   &  17.92 & 167.77  & 57.87\\
263.60  &  -1.57 & 192.39  &  30.87 & 17.83   &  19.73 & 69.46   &  29.80\\
240.96  &  23.13 & 57.24   &  26.95 & 269.34  &  74.10 & 198.98  &  53.16\\
297.94  &  18.57 & 323.63  &  7.87 & 247.29  &  34.70 & 213.73  &  37.93\\
294.35  &  70.98 & 293.83  &  -5.91 & 33.82   &  13.57 & 166.55  &  42.07\\
146.04  &  23.93 & 258.77  &  56.35 & 167.52  &  56.27 & 348.37  &  12.03\\
293.67  &  50.59 & 55.38   &  44.74 & 287.13  &  5.22 & 113.95  &  32.31\\
55.59   &  49.34 & 345.16  &  33.63 & 339.45  &  42.34 & 68.59   &   5.00\\
59.55   &  51.56 & 211.47  &  37.34 & & & & \\
\end{tabular}
\label{agasa}
\end{table}


\begin{thebibliography}{99}
\bibitem{ylm} Throughout this work 
we use real-valued spherical harmonics, which are obtained from the complex ones by substituting, 
$e^{i\,m\,\phi} \rightarrow \sqrt{2} \sin (m\phi)$, if $m<0$, $e^{i\,m\,\phi} \rightarrow \sqrt{2} 
\cos (m\phi)$, if $m>0$, and $e^{i\,m\,\phi} \rightarrow 1$ if $m=0.$
\end{thebibliography}
\end{document}